\documentclass[%
 reprint,
superscriptaddress,
 amsmath,amssymb,
 aps,
prb,
]{revtex4-2}
\usepackage{graphicx}
\usepackage{float}
\usepackage{epstopdf}
\usepackage{color}
\usepackage{amsmath}
\usepackage{amssymb}
\usepackage{wasysym}
\usepackage[colorlinks = true,
            linkcolor = blue,
            urlcolor  = blue,
            citecolor = blue,
            anchorcolor = blue]{hyperref}
\usepackage[pagewise,modulo]{lineno}

\usepackage{academicons}
\usepackage{tikz,xcolor,hyperref}

\definecolor{lime}{HTML}{A6CE39}
\DeclareRobustCommand{\orcidicon}{%
	\begin{tikzpicture}
	\draw[lime, fill=lime] (0,0) 
	circle [radius=0.16] 
	node[white] {{\fontfamily{qag}\selectfont \tiny ID}};
	\draw[white, fill=white] (-0.0625,0.095) 
	circle [radius=0.007];
	\end{tikzpicture}
	\hspace{-2mm}
}
\foreach \x in {A, ..., Z}{%
	\expandafter\xdef\csname orcid\x\endcsname{\noexpand\href{https://orcid.org/\csname orcidauthor\x\endcsname}{\noexpand\orcidicon}}
}

\usepackage{etoolbox}
\usepackage{tikz}

\newrobustcmd*{\mycircle}[1]{\tikz{\filldraw[draw=#1,fill=#1] (0,0) circle [radius=0.07cm];}}
\newrobustcmd*{\myholowcircle}[1]{\tikz{\filldraw[draw=#1,fill=white] (0,0) circle [radius=0.07cm];}}
\newrobustcmd*{\myholowsquare}[1]{\tikz{\filldraw[draw=#1,fill=white] (0,0) rectangle ++(4pt,4pt);}}
\newrobustcmd*{\mytriangle}[1]{\tikz{\filldraw[draw=#1,fill=#1] (0,0) --(4pt,0) -- (2pt,4pt);}}
\newrobustcmd*{\myholowtriangle}[1]{\tikz{\filldraw[draw=#1,fill=white] (0,0) --(4pt,0) -- (2pt,4pt);}}
\newrobustcmd*{\mydowntriangle}[1]{\tikz{\filldraw[draw=#1,fill=#1] (-2pt,4pt) --(0pt,0pt) -- (2pt,4pt);}}
\newrobustcmd*{\myholowdowntriangle}[1]{\tikz{\filldraw[draw=#1,fill=white] (-2pt,4pt) --(0pt,0pt) -- (2pt,4pt);}}

\usepackage{natbib}
\begin{document}

\title{Fast ion diffraction of protons on NaCl,\\ the discovery of GIFAD}

\author{Patrick Rousseau\orcidA{}}
\affiliation{Normandie Univ., ENSICAEN, UNICAEN, CEA, CNRS, CIMAP, Caen, 14000 France},
\author{Philippe Roncin\orcidB{}}
\affiliation{Universit\'{e} Paris-Saclay, CNRS, Institut des Sciences Mol\'{e}culaires d'Orsay (ISMO), Orsay, 91405 France}%91405


\begin{abstract}
Grazing incidence fast atom diffraction (GIFAD or FAD) has become a technique to track the surface topology
of crystal surface at the atomic scale. The paper retraces the events that led to the
discovery of unexpected quantum behavior of keV atoms during the thesis of Patrick Rousseau in Orsay and
Andreas Schueller in Berlin. In Orsay, it started by diffraction spots whereas in Berlin supernumerary rainbows were
first identified at keV. Though the discovery was not anticipated, it did not take place by accident, everything
was in place several years before, waiting only for an interest in neutral projectiles with a touch of curiosity\footnote{Presented at Int. conf. on Inelastic Ion-Surface Collisions (IISC-24) Charleston Sept. 2023 Accepted to NIMB Ed. Chad Sosolik}. 
\end{abstract}

\maketitle
%\tableofcontents
\newpage
%\begin{linenumbers}	
\section{Introduction}\label{ch:intro}
%% \linenumbers

%% main text

An unexpected outcome of the ion-surface community (IISC) was the discovery of grazing incidence fast atom diffraction, (GIFAD\cite{Rousseau_2007} or FAD\cite{Schuller_2007}).
This technique links the separate worlds of the keV energy range where inelastic atomic collisions prevail and the meV energy range discovered in the 1930s by I. Esterman and O. Stern\cite{estermann_1930} demonstrating the wave nature of atoms. 
These pioneers opened the field of atom surface interactions with the discovery of molecular rays\cite{knauer_1929reflexion} and the universal attractive forces\cite{frisch_1933anomalien,lennard_1936diffraction} that would develop into molecular beam epitaxy\cite{toennies_2011otto} (MBE), surface catalysis and surface science in general.
The connection between these two energy domains is the strong decoupling of the fast movement along the crystal axis where the primary beam is oriented and the much slower one in the perpendicular plane and this could be quantified by the obliquity factor derived by P. Kapitza and P. Dirac\cite{kapitza1933reflection} who predicted the diffraction of matter waves by standing light waves long before the discovery of lasers. 
Note that a number of authors have provided details or rediscovered numerically the effect as listed in Ref.\cite{Henkel}.
%See Ref.\cite{Henkel_94} for adaptation to atoms on evanescent waves above a dielectric surface, Ref.\cite{Debiossac_PRA_2014} for adaptation to GIFAD, Refs.\cite{brandt1968channeling,krause1986rainbow} for the channeling of MeV ions in crystals and Refs.\cite{danailov_1991scattering,danailov2001deduction,farias_2004_pronounced,zugarramurdi_2012_transition,muzas_2016_diffraction} for independent rediscoveries of the decoupling effect also named axial surface channeling approximation (ASCA).
The large projectile energy of GIFAD confines the entire diffraction pattern into a narrow cone that can be imaged at once onto a position-sensitive detector (see \textit{e.g.} Ref\cite{pan2022setup} for a GIFAD setup).
GIFAD can track online the growth of successive layers by the oscillation of the scattered intensity\cite{Atkinson_2014} as was already established with keV ions in the IISC community\cite{fujii1993layer,igel1996intensity}.
The advantage of using atoms is that the diffraction pattern of a layer may be rich enough to provide a perfect fingerprint of the exact topology of the surface reconstruction\cite{Debiossac_PRB_2014} and that atoms are not sensitive to electromagnetic fields.
Also, the impact energy is not limited by image charge effects ($\approx$1 eV) and the absence of charging makes GIFAD ideal for monitoring the growth of fragile organic layers\cite{Seifert_2013, Kalashnyk_2016}.

The paper is organized as follows.
Section \ref{ch:status} recalls the scientific context in two separate communities, that of inelastic ion scattering at surfaces (IISC) and that of highly charged ions (HCI) in the early 1990s and the significant merging that took place in a few years, with a special emphasis to the grazing incidence geometry.
Section \ref{ch:NaCl} presents results on the charge exchange and energy loss process taking place during the collision of H$^+$ ions on the NaCl(001) surface, the system that led to the first diffraction images of fast ions and fast atoms diffraction on surfaces. Section \ref{ch:FAD_Berlin} presents the discovery of FAD in Berlin before Section \ref{Ch:competition} draws some aspects of the intense competition that took place between Orsay and Berlin.

\section{IISC and HCI communities at the turn of the century}\label{ch:status}

This section recalls a personal view of the scientific context in the communities of highly charged ions (HCI) and of ion-surface collisions (IISC) that were initially only weakly connected by a common interest in atomic collisions.
Both communities were partly funded by basic science programs toward future fusion devices.
Highly charged ion impurities had been identified by spectroscopic techniques and were considered responsible for a significant power loss due to X-ray emission after electron capture collisions from the injected H$^\circ$ and D$^\circ$ atoms to be used both as a fuel for the nuclear reaction and a heat source to reach the required plasma temperatures.
%With who were worried by power loss due to X-ray emission after electron capture collisions with impurities in the plasma and by material damage due to atomic collisions on the walls.
In this respect, a major concern was to increase the energy of these H$^\circ$ atoms injected in the plasma to MeVs thereby forcing the community to abandon the production of H$^\circ$ by electron capture from H$^+$ ions and to develop a device based on electron detachment from H$^-$ ions.
Any scientific and technological improvement in the production of intense H$^-$ beams was strongly encouraged and financed. %(W sputtering, hydrogen retention etc..material damage due to energetic ion etc..)
In Europe, the FOM Amsterdam\cite{geerlings1985h,los1990charge} invested large theoretical and experimental efforts to lower the surface work-function of metal surfaces with Cs deposition.
Powerful dynamical rate equations describing the transitions back and forth between the projectile and the surface were developed.
These outlined the role of the "way out" of the collision when the projectile leaves the surface and where the attenuation of loss rate to the metal would eventually decide for the final H$^-$ production\cite{geerlings1985h,los1990charge,borisov_1992h,maazouz1996h,borisov1998finite}.
%These research pioneered in the 80's at FOM Amsterdam\cite{geerlings1985h,los1990charge}  have reached impressive  achievements in the 90's combining experiment and theory.

In the early 90's many ion-surface experiments were trying to isolate quasi-head-on collisions with surface and sub-surface atoms at large angles of incidence\cite{niehus1993low}.
Analytic and structural applications had developed around the concepts of shadow cone\cite{oen1983shadow} that are at the heart of many experimental techniques TOF-SIMS\cite{delcorte1997tof}, LEIS\cite{Taglauer_1993leis}, TOF-SARS\cite{grizzi1990tofsars}, DRS\cite{masson1989application}.
Most of the inelastic effects were considered from the electronic point of view and associated with inner-shell excitation\cite{esaulov1994autoionising} and a significant penetration below the surface.

Hyperthermal ions (1$\lessapprox$ E$\lessapprox$ 50 eV) are much more surface-sensitive and the progressive evolution of the single binary collision condition to a multiple collision regime is well illustrated in the work B.H. Cooper.
In\cite{dirubio1996energy}, the quasi-single, quasi-double, and quasi-triple collision peaks of Na$^+$ ions on Cu(001) surfaces are identified as separate peaks in the energy loss spectra.
The progressive merging of these peaks to an unresolved broad structure as the energy is decreased or as the incidence angle becomes more grazing was clearly observed.
The comparison with trajectory simulation outlined the role of the image charge attraction on the effective incidence angle and the role of thermal motion of the surface atoms when these quasi-single, double, and triple collision processes become hardly distinguishable\cite{johnson2022safari}.  
A good control of the kinematics has also promoted detailed investigations of the charge exchange processes taking place at the surface (see also the work of J.A.Yarmoff).
%In general, the regime of multiple scattering was considered as unable of providing 

\subsection{Grazing scattering of ions on metal surfaces}\label{ch:grazing_status}
Surprisingly, from today's perspective, the ion-surface grazing scattering geometry was developed in nuclear physics as a means to produce polarized nuclei at MeV energies\cite{rau1973measurement,winter1979nuclear}.
It evolved to keV energies focusing on the ion-surface interactions under well controlled distances to the surface revealing contributions of the band structure in metals\cite{winter1983band}.
The technique was developed with lower primary kinetic energy providing more gentle interaction with the surface, thereby improving our understanding of the decoupling of the fast motion parallel to the surface from the slower one in the perpendicular plane.
The jellium model\cite{gadzuk1983vibrational} was developed using an ideal flat metal surface as the support of an electron sea characterized by its electron density and specific decay range at the surface.
This triggered several original contributions to the physics of negative ion production\cite{wyputta1991h} and to the measurements on image charge acceleration\cite{winter1992image,winter1993image}.
The description of the atomic levels and lifetime in front of a surface significantly improved\cite{borisov_1993energy} allowing a fresh view of basic atomic processes above the surface such as the Auger-Meitner transition rate\cite{cazalilla1998theory,hecht_1998auger}, the triplet to singlet conversion of metastable helium\cite{borisov_1993singlet} via electron exchange with the surface.
It also pushed a more refined description of the surface emphasizing the role of surface states\cite{maazouz1998electron} and of the projected band-gap\cite{hecht_2000role} as well as new models such as the shifted Fermi sphere\cite{borisov_1996shifted} to account for the dependence with the velocity parallel to the surface\cite{danailov_1991scattering}.
 
Once inside the metal, oscillations of the energy loss to the electron gas with the nuclear charge could be observed and interpreted in terms of effective electron density in the conduction band\cite{juaristi1999charge}.

A few experiments started to investigate the surface channeling conditions with position-sensitive detectors in Kyoto\cite{fujii1988surface} and Osnabrück\cite{niehof1990experimental} to image the specific scattering profiles that were only qualitatively understood.

\subsection{The arrival/merging of the highly charged ion community}
The HCI community gathered in the mid 1980s when new generations of ion sources were starting to produce HCI at low kinetic energy.
HCI were already available in the MeV energy range behind stripping foils and a few atomic physicists were investigating the large number of photons emitted beyond the foil\cite{Martin_1985} as well as high energy collisions in gas jets\cite{ULLRICH_1986,STOLTERFOHT_1971} or the channeling of fast ions through crystals aligned along low index directions\cite{krause1986rainbow}.

These new ion sources were the electron beam ion source\cite{donets1998historical} (EBIS) originally developed in nuclear physics facilities to bypass the first acceleration + stripping stage of all high energy devices at this time.
Physicists already present around the accelerator were the first to use these ions sources \cite{iwai1982cross,briand1984observation}.
However, though rather selective in charge state EBIS had a limited capacity 
and electron cyclotron resonance (ECR) sources\cite{geller2018electron} which were byproducts of the research on fusion devices demonstrated much larger ion current.

Fundamental investigations were sponsored by the fusion department interested in a spectroscopic diagnostic tool\cite{hoekstra1990state}, as well as cross sections\cite{muller1977scaling,iwai1982cross} combined with general schemes of single and multiple electron capture\cite{barat1992multiple}, electron impact excitation\cite{kilgus1992high} to feed the plasma modeling.

The capacity of HCI to capture large numbers of electrons into excited states formed of hundreds to thousands of states, and the subsequent atomic decay triggered intense theoretical and experimental developments, new electron and photon spectrometers, new detectors to count secondary electrons\cite{lakits1990threshold} etc.
Following a suggestion from D. de Bruyn from FOM Amsterdam, I assembled the first 2D position sensitive detector (PSD) based on a backgammon anode combining electrical and geometrical charge division technique to be placed in an electrostatic analyzer\cite{roncin_1986}.
The associated ability to record simultaneously a large range of energy loss and scattering angle offered a significant advantage over competitors using energy loss spectroscopy in Belfast, Stockholm, Aarhus, Bochum, Vienna, Tokyo, Nagoya, and in Kansas which sometimes had better energy resolution\cite{AIM}. 
Most often, the instruments were designed to be operated in coincidence so that the spectra could be correlated, for instance the scattering profile with the emitted photon polarization to determine in which sense the captured electron was rotating\cite{roncin1990observation}, if specific X-rays were emitted from different atomic doubly excited $(nl,n'l')$ manifolds\cite{chetioui1990doubly}, or if visible light associated with Rydberg could be associated with a specific primary collisional process\cite{Bernard_1997}.
Note that the development of PSD played a key role in the advent of recoil ion momentum spectroscopy\cite{moshammer_1996_4pi,flechard1997state} defining a new standard in "collision spectroscopy" in the late 1990's.

After a few years investigating the primary collisional processes at play in electron capture in the gas phase, the HCI community started to specialize and evolve in new directions: QED in atoms\cite{indelicato1992relativistic}, optical metrology\cite{king2022optical}, mass spectrometry\cite{Rousseau_2010}, large\cite{thomas2011double} and small\cite{bernard2008tabletop} storage devices, interaction with clusters\cite{manil2003highly} and in surface modifications after the demonstration that a single HCI impacting a surface can emit hundreds of electrons\cite{aumayr1993emission}.

In 1993, a dedicated program of the European Union coordinated by N. Stolterfoht provided funds to favor "human mobility" between European labs with a special focus on surfaces\cite{arnau1997}.
One of the first decisions was to purchase a set of LiF(100) single crystals for each participating group because material modification was expected to be more important on insulators while charging up can be avoided by heating up to a few hundred of $^\circ$C to favor ionic mobility at the surface.
\begin{figure}\centering
\includegraphics[width=0.95\linewidth,draft = false]{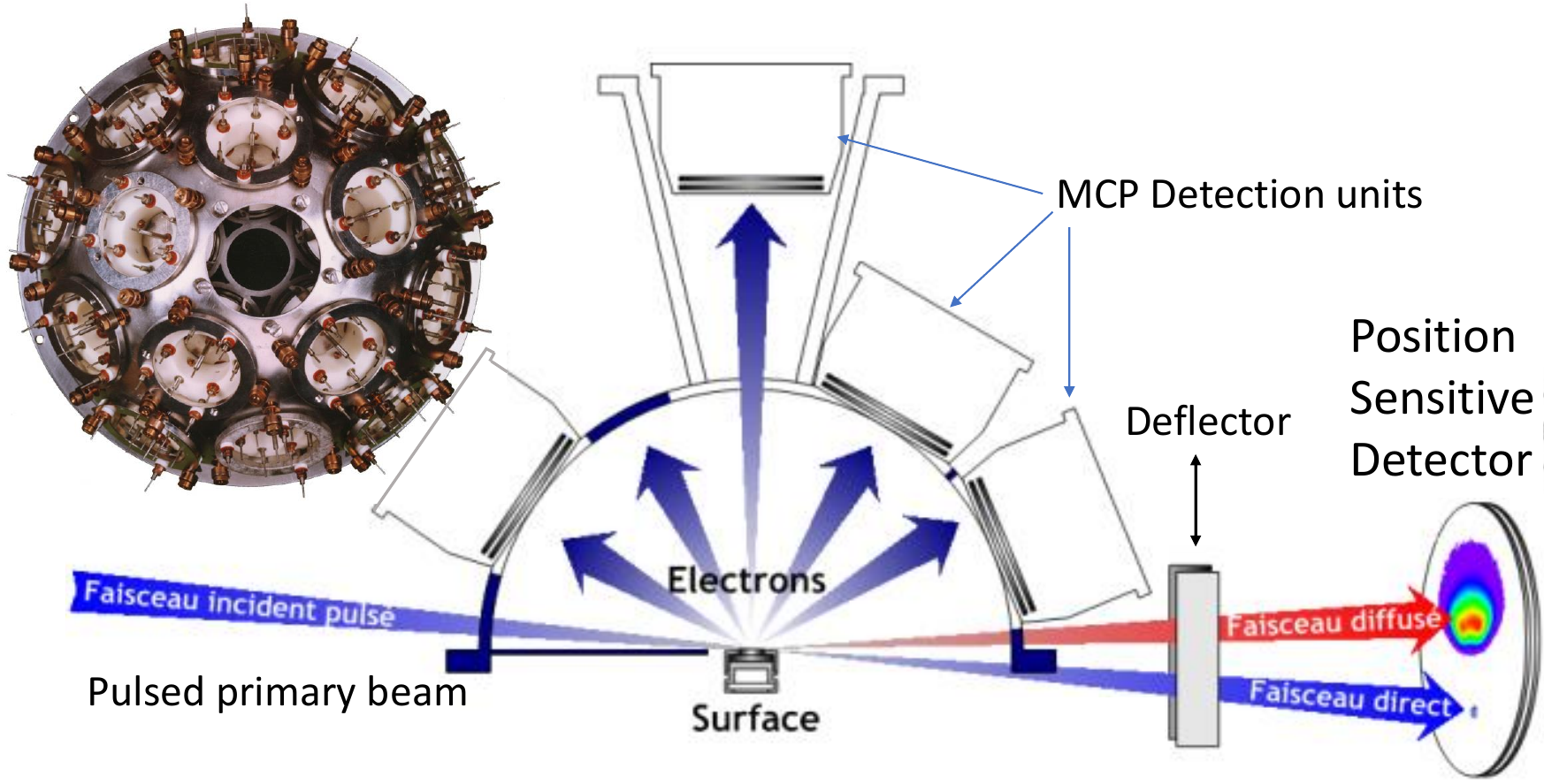}
\caption{\label{fgr:Setup} Sketch of the experiment designed to measure in coincidence, emitted electrons and scattered projectile analyzed on a position and time-sensitive detector. Electrons and projectile energies are measured by time of flight referred to the electrostatic chopper. The inset is a picture of the 2$\pi$ detector designed by V.A. Morosov\cite{Morosov_1996}.}
\end{figure}
I was dreaming of a multi-detector setup and M. Barat presented the project to Pr Leonas from IPM Moscow who encouraged a new collaboration.  
Together with V.A. Morosov, A. Kalinin and Z. Szilagyi, we started to build the multi-detector sketched in Fig.\ref{fgr:Setup}. It is made of 16 sub-detectors able to detect secondary electrons or ions together with a position-sensitive detector for the scattered projectile\cite{Morosov_1996}.
With a pulsed primary beam\cite{Chopper}, the energy of all the products could be analyzed by time of flight.

\subsection{Grazing scattering on insulator surfaces}\label{ch:PSD}
The movement also attracted members of the IISC community who were the first to measure the image charge acceleration of these ions towards the surface under grazing incidence\cite{auth1995image}.
The Berlin group had switched to insulators with remarkable success. 
Following a proposal by F.J. Garcia de Abajo and A.G. Borisov at the IISC-11 conference in Wangerooge, they observed resonant coherent excitation of the Lyman $\alpha$ line of H$^\circ$ atoms flying over the electric field above the LiF surface\cite{auth_1997resonant}.
They could observe complete negative ion conversion of positively charged halogen ions\cite{winter1996complete,MB_negative_ions} and derive a general and powerful electron capture scheme by realizing that, close to the surface, the projectile shares almost the same Madelung potential as the surface ions so that the "energy defect" is locally reduced on top of the halogen site\cite{auth1995high,borisov_1996_diabatic}.

Then they observed a threshold in the velocity dependence of the energy loss of H$^\circ$ on LiF\cite{auth1998threshold}.
This proved that the surface of ionic insulators is not as full of electronic defects as was commonly assumed from the unexplained large secondary electron yield, a topic dear to the Vienna group\cite{winter1991recent,khemliche2009electron}.

%On our side, the interaction of highly charged ions with surfaces turned out to more  complex than anticipated. 
On our side, investigating HCI-surface with our multi-detector did not provide the expected results. 
At grazing incidence, whatever the initial charge state the dominant products were neutrals and negative ions\cite{auth1995high,MB_negative_ions} with a large amount of emitted electrons but with an unexpectedly large energy loss\cite{roncin1999_Oq} hardly compatible with quasi-resonant capture models\cite{Ne2}.
We measured correlations between all parameters, energy loss, scattering angle, and number of electrons but we could not improve the general description.
This is in part because the simplest things do not always come first. 
It was only a few years after that the Berlin group could measure tens of eV energy loss of scattered Ne$^+$ ions that could be interpreted quantitatively as the excitation of optical phonons\cite{borisov1999evidence}.
As a positive ion flies over an ionic crystal, it attracts the halogen ions and repels the alkali by coulombic potential, and excites optical phonon on the surface producing a velocity and $z$-dependent friction force.

We decided to reduce the complexity by investigating single electron capture processes by protons. 
With J. Villette, we measured the correlation between energy loss and electron emission with a mean value of 34 eV per emitted electron\cite{villette_1999grazing} whereas the energy needed to extract an electron from the valence band is around 13 eV.
We did not even consider using neutral H$^\circ$ projectiles (if we had, we would probably have discovered GIFAD) but we could identify an energy loss peak that is not associated with electron emission that instead corresponds to a surface electronic excitation identified as the surface exciton\cite{Roncin_1999}.
With A.G. Borisov we could relate all observations to a primary formation of H$^-$ as elaborated earlier in Berlin\cite{borisov_1996_diabatic} and Orsay\cite{borisov1997theory} that we could enrich by introducing the quasi-molecular curve crossing with the exciton levels and by an electron detachment process when the H$^-$ ion flies over a surface F$^-$ ion explaining altogether, the large secondary electron emission, the large energy loss and the moderate fraction of negative ions\cite{Roncin_1999}.

With J.Villette we started to measure the energy loss of neutral projectiles but we chose Ne atoms (here again the choice of helium could have led to the observation of diffraction).
We observed the first log-normal scattering profile and clear signs of a $E\theta^3$ 5$E$ the primary energy and $\theta$ the angle of incidence) dependence of the energy loss\cite{Villette_these}, two processes that would further be connected to inelastic diffraction\cite{roncin2017elastic}.
However, the Berlin group was the first to publish comparable results \cite{mertens2000energy} and we never published.

Using the coupling to optical phonons as a friction force we could identify a skipping motion above the surface due to the trapping of Ne$^+$ ions by their own image charge\cite{Villette_2000,snowdon1988observation} while the slow neutralization at each bounce could be attributed to "dark" Inter-Atomic-Auger-Meitner decay where one electron is captured from a F$^-$ ion while the one of an adjacent F$^-$ ion is excited to form an electron-bi-hole state known as a trion\cite{khemliche2001electron}.

We started to be rewarded for our hard work on the $2\pi$ detector and I was proud to offer a collaboration to the Berlin group on a double electron capture mechanism. 
They found that on LiF, primary F$^+$ ions produce more F$^-$ than primary F$^\circ$ while we had a similar result with fewer data points but resolved in energy loss and electron emission\cite{roncin_2002_F}.
With A. Borisov we could show that both electrons are captured from the adjacent sites in a correlated way but due to two independent single electron processes forced to act together for energetic reasons, a process similar to the one that I investigated with M. Barat in the gas phase collisions of C$^{4+}$ ions on helium\cite{barat1990single,roncin1987transfer,roncin2020revisiting}.
Here also, most often, the captured electron was found to be "recaptured" into the local excited states, a surface trion\cite{borisov2003f}.
The collaboration was a success but Helmut Winter was very clear that this should never happen again.
There was no animosity of any sort but he considered that competition was the best ingredient to keep the field of grazing collisions alive.

\section{The H$^+$ NaCl system} \label{ch:NaCl}
\subsection{Experimental setup}
\begin{figure}\centering
\includegraphics[width=0.8\linewidth,draft = false]{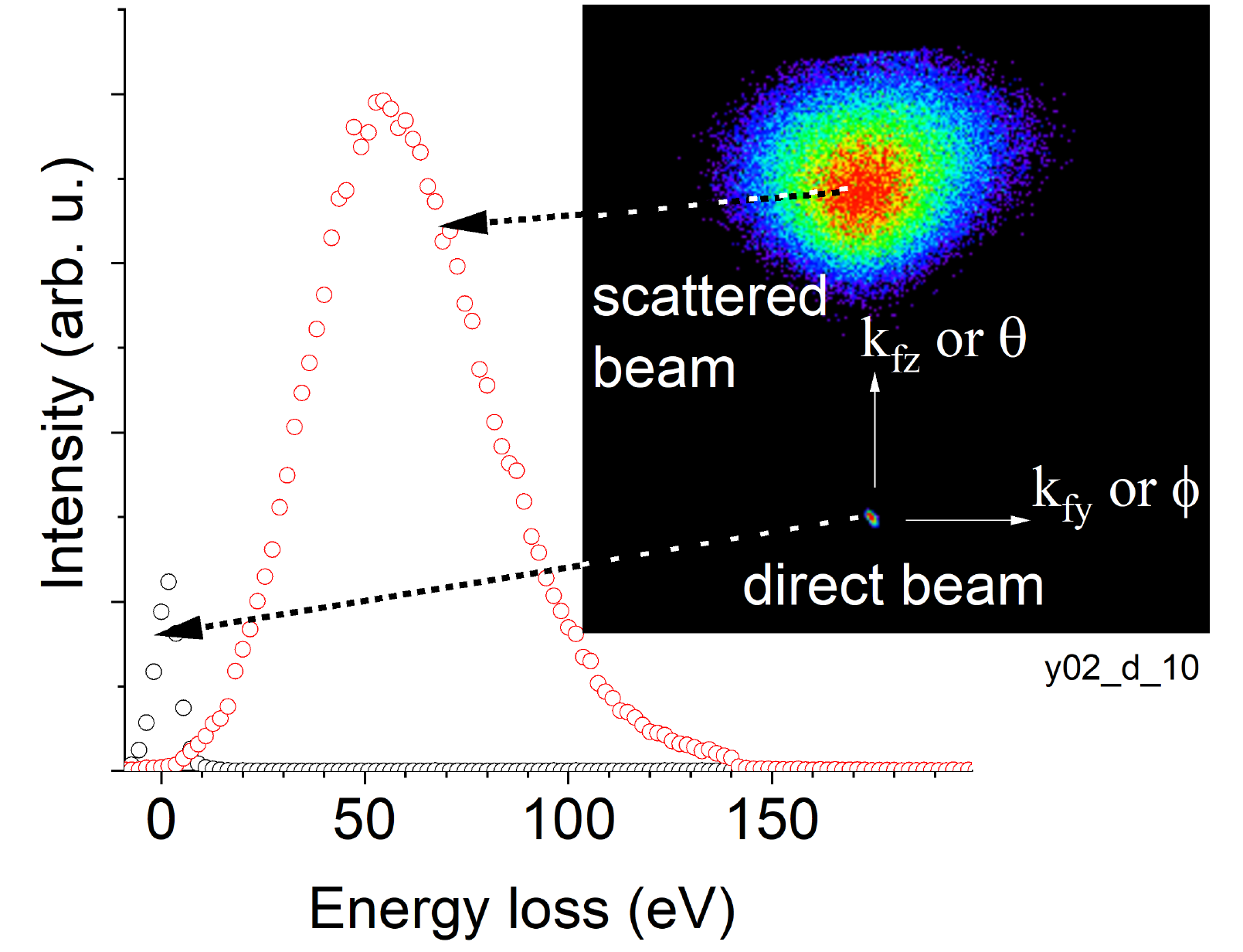}
\caption{\label{fgr:y02_d_10} For 1 keV H$^+$ ion on NaCl(001) along a so called random direction [Rnd] with $\theta_i$=1.6$^\circ$. The inset is a PSD image showing the broad quasi-symmetric ($\sigma_\theta\sim\sigma_\phi$) scattering profile. The tiny spot labeled "direct beam" is present when the target surface is not fully inserted. The associated energy loss spectra peak at $\overline{\Delta E}\sim$ 60 eV. (Taken from Ref.\cite{Rousseau_these})}
\end{figure}

The experimental setup has been described in Ref.\cite{Morosov_1996} and is sketched in Fig.\ref{fgr:Setup}.
Briefly, a beam of 300 to 1000 eV H$^+$ ions is collimated and chopped by electric pulses applied to deflection plates before entering the UHV chamber via another tiny aperture. 
It interacts at grazing incidence $\theta_i\sim0.5-3^\circ$ with a crystal surface placed in the center of a $2\pi$ electron detector and the scattered particles are detected onto a PSD placed 30 to 50 cm downstream.
The beam direction and the normal to the surface ($z$) define the ($x,z$) collision plane and the incident angles $\theta_i$ and azimuth $\phi_i=0$.
Within minor corrections, the impact location on the PSD is a direct measure of the scattering angle ($\phi_f,\theta_f$) or final momentum ($\hbar k_{fy},\hbar k_{fz}$). 
The scattered particle charge state can be analyzed by inserting two deflection plates at the exit of the $2\pi$ detector.
Usually, the target surface is not fully inserted in the beam so that a fraction of the beam hits the detector providing a reference for the time of flight as illustrated in Fig.\ref{fgr:y02_d_10}.
The energy resolution is measured as the standard deviation $\sigma_b$ of the direct beam and can be close to 1 eV (see Table I).%\ref{Tab:ty02_d_16}).
In addition, the absolute positioning relative to the direct beam is sometimes affected by an uncertainty close to 1 eV, measured as a slight difference between the ions created when the pulsing voltage goes up or when it goes down.
%This does not affect the relative values \textit{i.e.} the distance between peaks.

Each impact corresponds to a time of flight $t$, a lateral $\phi_f$, and a polar $\theta_f$ deflection angle, so that each pixel in the 2D images is a time of flight spectrum (making a 3D array). 
More precisely, each 2D image displayed here is a particular slice of the 3D ($k_{fy}, k_{fz},t$) corresponding to selected values of the time of flight between $t_{min}$ and  $t_{max}$.
In addition, each 3D array/graph can be restricted to specific conditions related to the number of secondary electrons detected, their direction, and their time of flight.
The distributions recorded in coincidence with a given number of detected electrons can be transformed into distributions correlated with a given number of emitted electrons by taking into account the absolute detection efficiency measured \textit{in situ} with an $^{241}$Am source\cite{villette_2000_cal}.
%The table \ref{Tab:NaCl} report typical energies measured on NaCl when investigated electron capture by C$^+$, F$^+$ Ne$^+$, Ar$^+$ and Kr$^+$ ions under grazing incidence angle\cite{rousseau2007auger,Rousseau_these}.

%\subsection{results}
%\begin{table}
%\centering
%\begin{tabular}{|p{2.2cm}|p{1.2cm}|p{1.2cm}|p{1.2cm}|}
%\hline
% E in eV&$E_{bv}$ & $E_{exciton}$&$E_{trion}$\\
%\hline
% LiF\cite{Roncin_1999,khemliche2001electron}&13$\pm$1& 1.1$\pm$0.5&3.5$\pm$1\\
% NaCl\cite{rousseau2007auger,Rousseau_these} &10.3$\pm$1& 1.6$\pm$1&3.1$\pm$1\\
% \hline
%\end{tabular}\label{Tab:NaCl}
%\caption{For LiF and NaCl surfaces, $E_{bv}$ is the energy needed to extract an electron from the valence band, $E_{exciton}$ and $E_{trion}$ are the binding energy of an excited electron bound to a single F$^\circ$ or Cl$^\circ$ hole and energies or two adjacent holes respectively.}
%\end{table}

When exploring a new system, the first experiment is usually to determine how reactive it is.
Fig.\ref{fgr:y02_d_10} shows the full scattering profile of 1 keV H$^+$ on NaCl along a direction that is far from a low index crystal axis. Since this is usually the case if no careful alignment has been prepared these are traditionally called random direction and indicated hereafter as [Rnd].
The mean energy loss of 60 eV indicates that, in addition to the electron needed for quasi-resonant neutralization, around six additional electrons are removed from the valence band while the electron detectors indicate that almost two electrons are emitted which is slightly more than from LiF in comparable conditions\cite{villette_1999grazing}.
The location, width, and relative intensities of all peaks associated with scattered H$^+$, H$^\circ$, and H$^-$ are reported in Table I.%\ref{Tab:ty02_d_16}).

\begin{figure}\centering
\includegraphics[width=0.8\linewidth,draft = false]{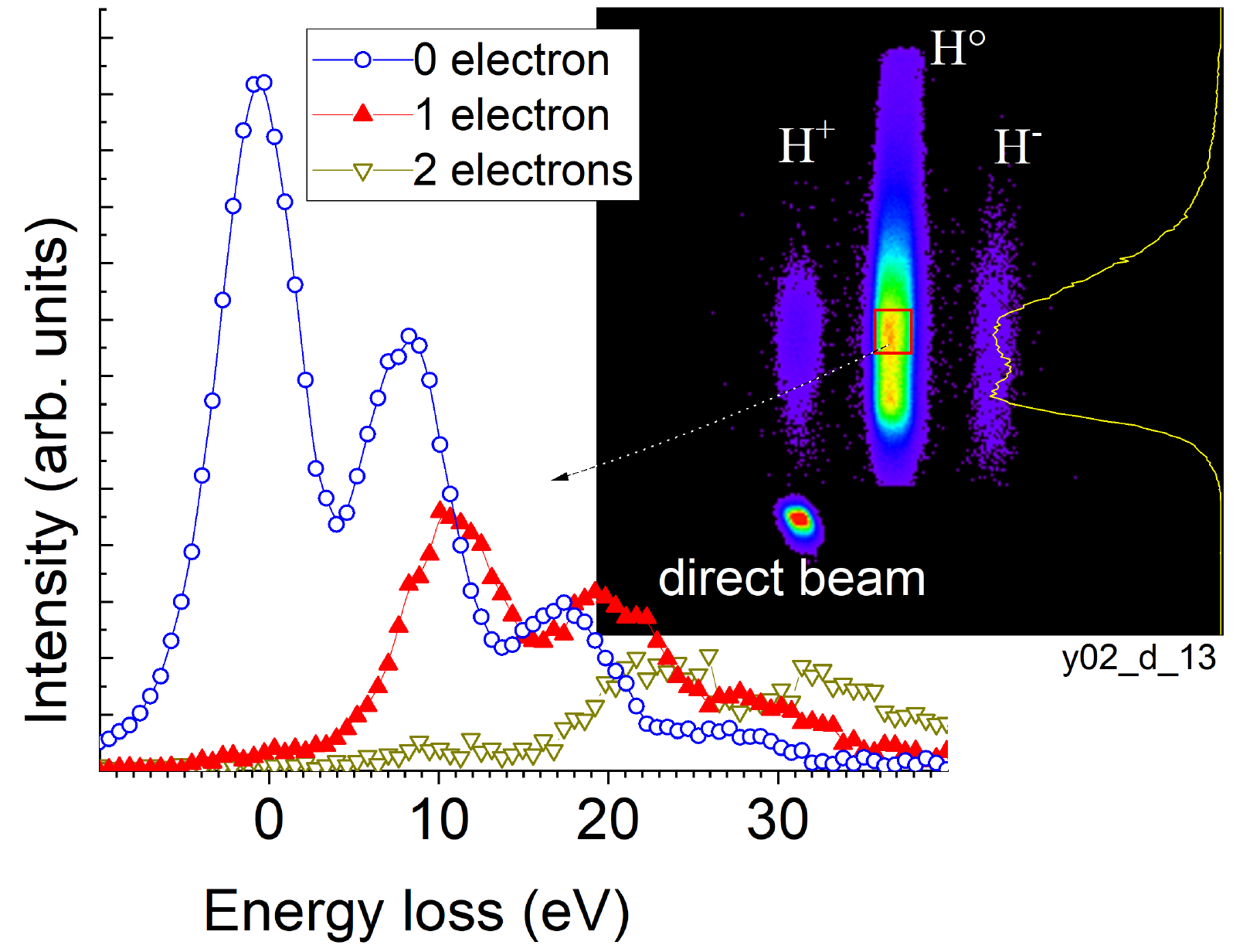}
\caption{\label{fgr:y02_d_13} For 500 eV H$^+$ ion on NaCl along [Rnd] with $\theta_i \sim$0.7$^\circ$. The inset is an image of the scattering profile after passing through a slit and deflection plates. The yellow line on the right is the polar profile of the central band corresponding to scattered H$^\circ$. The energy loss spectra corresponding to the zone in red on the inset are plotted as a function of the number of emitted electrons.(Taken from Ref.\cite{Rousseau_these})}
\end{figure}

\begin{table}
\centering
\begin{tabular}{ |p{1.5cm}|p{1.1cm}|p{1.1cm}|p{1.3cm}| }
\hline
    &$\Delta E (eV)$& $\sigma_E (eV)$&ratio (\%)\\
\hline
%direct & 0 & 1.5  & \\
H$^+$  & +2.4  & 1.9  &8.8\\
H$^\circ$ 1$^{st}$& - 0.2 & 1.8  & 41.7\\
H$^\circ$ 2$^{nd}$& +7.8 & 2.7  & 39.3\\
H$^\circ$ + 1 e$^-$& +11.2 & 2.4  & 9.4\\
H$^-$ & +12.0 & 3.2  & 0.7\\
 \hline
\end{tabular}\label{Tab:ty02_d_16}
\caption{Position, width, and relative intensities of the peaks in the energy loss spectra displayed in Fig.\ref{fgr:y02_d_16}}
\end{table}

We started to lower both the energy and the angle of incidence by a factor $\sim 2$ thereby reducing the impact energy $E_\perp=E_0 \sin^2\theta_i$ by a factor $\sim 8$.
Fig.\ref{fgr:y02_d_13} shows the scattering profile of 500 eV H$^+$ through a slit and deflection plate showing three images of the slit associated with different final charge states: H$^+$, H$^\circ$, and H$^-$. 
The polar profile of the scattered H$^\circ$ is reported in yellow on the right-hand side of the insert and shows two peaks, a narrower one at a low exit angle and a broader one above.
The energy loss spectra reported in Fig.\ref{fgr:y02_d_13} correspond to a zone located on the broad maximum and indicated by a red rectangle.
The energy loss spectra correlated with 0, 1, and 2 emitted electrons are now well resolved with several peaks corresponding to a given number of electrons removed from the valence band similar to the spectra recorded on LiF at slightly larger impact energy\cite{Roncin_1999}. 
%\textcolor{red}{(0e :E0=0.5,dE=8.86 eV, sigmaE=2.56 dwE=0.61, 1e : E0=11.9, dE=9.17,we=2.9,dwE=1.29 (wEdb=2.043).}
The energy loss spectra corresponding to the scattering peak at a lower scattering angle are not shown but are comparable with almost equal intensity for the first peak but reduced intensities for the second and third peaks.

The scattered H$^+$ (not shown) loose only 1.5 eV indicating that these H$^+$ ions survived the interaction with the surface without capturing a single electron from the valence band.
The weak energy loss most likely originates from a weak coupling to the optical phonons\cite{borisov1999evidence}.

%\definecolor{OliveGreen}{rgb}#CDCD0D%{#CD,#CD,#0D}
\begin{figure}\centering
\includegraphics[width=0.8\linewidth,draft = false]{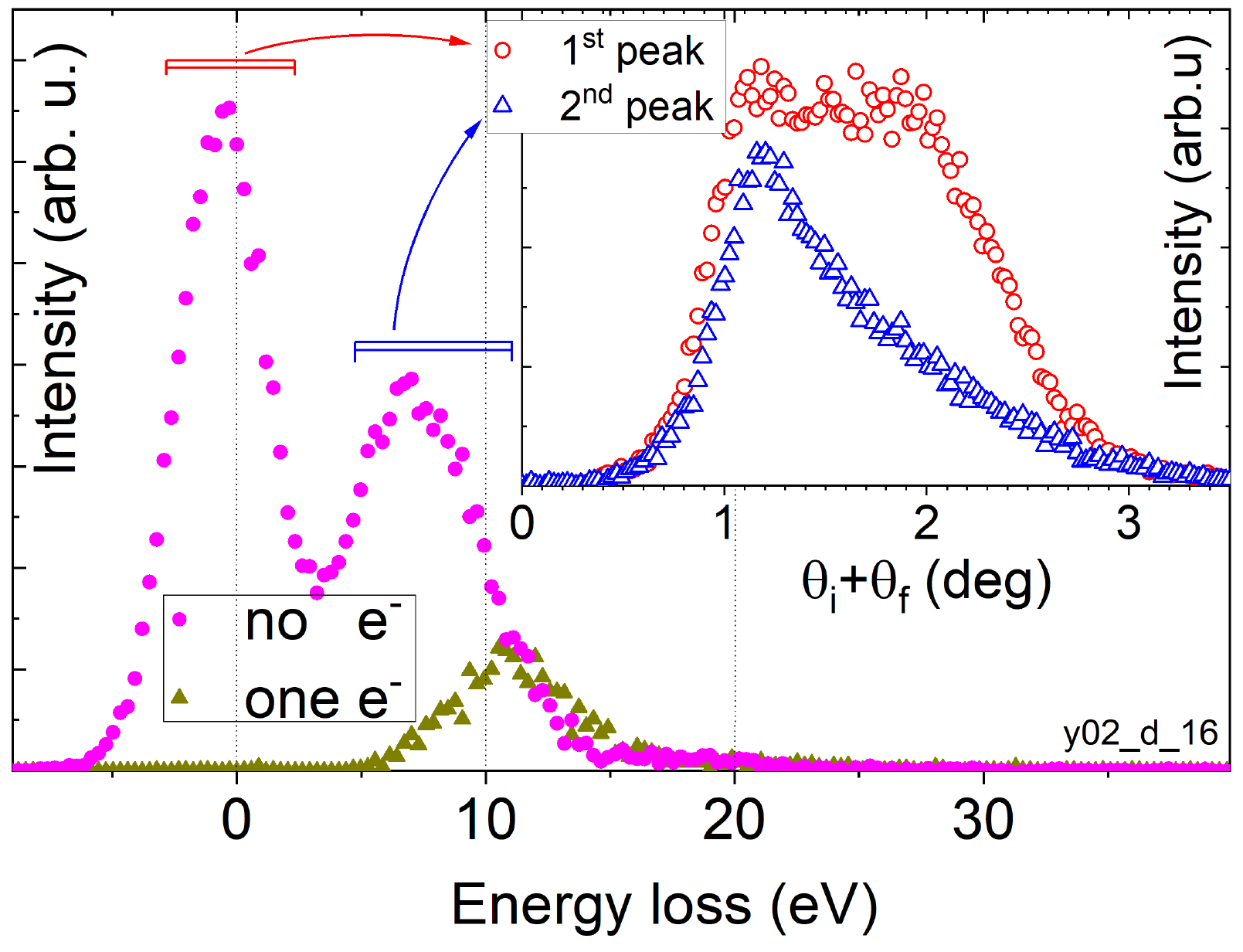}
\caption{\label{fgr:y02_d_16} The energy loss spectra of scattered H$^\circ$ correlated with the emission of one electron (\mytriangle{brown}) or with no electron emission (\mycircle{magenta}). The inset displays the polar scattering profiles associated with the first (\myholowcircle{red}) and the second peak (\myholowtriangle{blue}) of the energy loss spectrum irrespective of the number of emitted electrons. (Taken from Ref.\cite{Rousseau_these})}
\end{figure}

Reducing further the energy and angle to lower the number of peaks and their analysis, Fig.\ref{fgr:y02_d_16} shows only two peaks in the energy loss spectrum of H$^\circ$.
The polar scattering profiles $P(\theta)$ associated with the first and second peaks respectively are displayed in the insert. 
Both contribute to the peak at a low scattering angle and the first one contributes most at a larger scattering angle. 
This is different from the situation generally encountered with other systems where the more electrons removed from the valence band the larger the scattering angle. 

The first peak in the energy loss spectrum corresponds to the quasi-resonant neutralization of the H$^+$ ions.
In H$^+$+LiF the successive peaks could be interpreted as the formation, at further sites, of H$^-$ ions, most often immediately transferred to excitons states when crossing the associated levels.
The competing mechanism could also be a direct double electron capture as observed with the F$^+$+LiF system.
Both processes differ by the location of the two Cl$^\circ$ holes left on the surface. 
In the first case, they should be randomly distributed and mainly without mutual interaction.
In the second case, they need to be adjacent with a specific overall energy for their mutual repulsion and the increased binding energy of the excited electron to form a trion.
In terms of energy loss, the difference should correspond to $E_{trion}$ - $E_{exciton}$ \textit{i.e.} less than 2 eV, and the measured value is not fully conclusive.
The first process is obviously at work here because Fig.\ref{fgr:y02_d_13} indicates the presence of a third and fourth peaks, in other words, electron capture continues after the second peak while double electron capture should operate only once.
Still, the angular behavior suggests that the second peak is not simply the first one plus an additional excitation.
%Anouchah Momeni defended his PhD on 2003 January 6$^{th}$ about a week before I asked Patrick to investigate neutral hydrogene on NaCl!
In January 2003 after A. Momeni defended his thesis\cite{Momeni_these} and left the group, we reviewed with Patrick Rousseau all the systems investigated on the NaCl surface and I insisted that complementary data were needed on this system but using neutral H$^\circ$ projectile to disentangle the peculiar angular behavior.
I was convinced that the anomaly observed with H$^+$ ions was due to the part of the trajectory taking place after neutralization.
H. Khemliche was opposed because he thought that this was just a "trajectory effect" whereas, from the atomic collision point of view, a contrasted angular behavior is a genuine godsend.
I had to leave for a visiting committee but Hocine agreed to assist Patrick in operating the neutralization cell I developed with J. Villette during his thesis\cite{Villette_these}.
%They should be randomly distributed in the first case while they need to be adjacent in the second case.
\begin{figure}\centering\includegraphics[width=0.8\linewidth,draft = false]{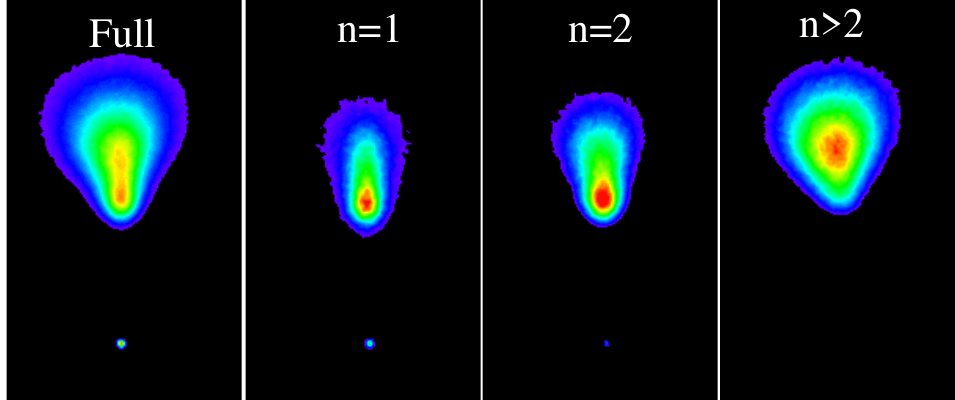}
\caption{\label{fgr:y03_a_03} The full 2D scattering profiles of H$^\circ$ atoms after impact of H$^+$ ions. On the left, no restriction is applied to the energy loss while the others are split according to the number n of electrons removed from the valence band. (see also Fig.3.44 of Ref.\cite{Rousseau_these} for comparable image along the $\langle 100 \rangle$ direction)}\end{figure}

\begin{figure}
\centering
\includegraphics[width=0.8\linewidth,draft = false]{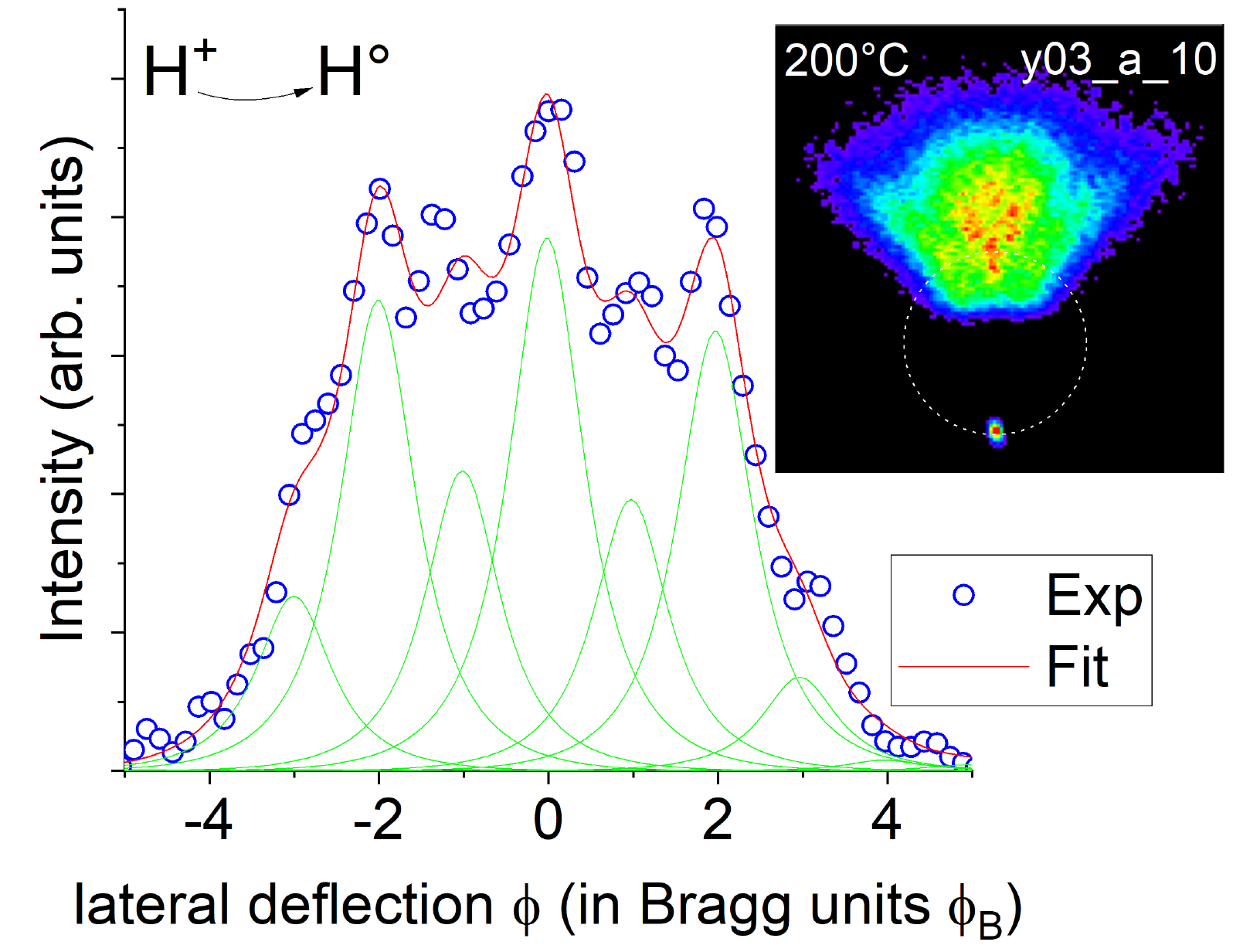}
\caption{\label{fgr:y03_a_10} First diffraction pattern of 500 eV H$^+$ projectiles on NaCl(001) along $\langle 100 \rangle$. Essentially quasi-elastic scattered H$^\circ$ are selected by restricting only the first peak in the energy loss. The angle of incidence is $\theta_i\sim$0.6$^\circ$ and the Bragg angle $\phi_B$=0.146$^\circ$. }
\end{figure}
The first experiment with H$^+$ ions on NaCl[Rnd] displayed in Fig.\ref{fgr:y03_a_03} confirmed the anomaly of the double-peaked angular structure with the two first peaks in the energy loss spectrum associated with the low scattering angle.
Changing the crystal orientation confirmed the peculiar angular behavior and Fig.\ref{fgr:y03_a_10} even shows early signs of inelastic diffraction with an ionic projectile.

\begin{figure}\centering\includegraphics[width=0.8\linewidth,draft = false]{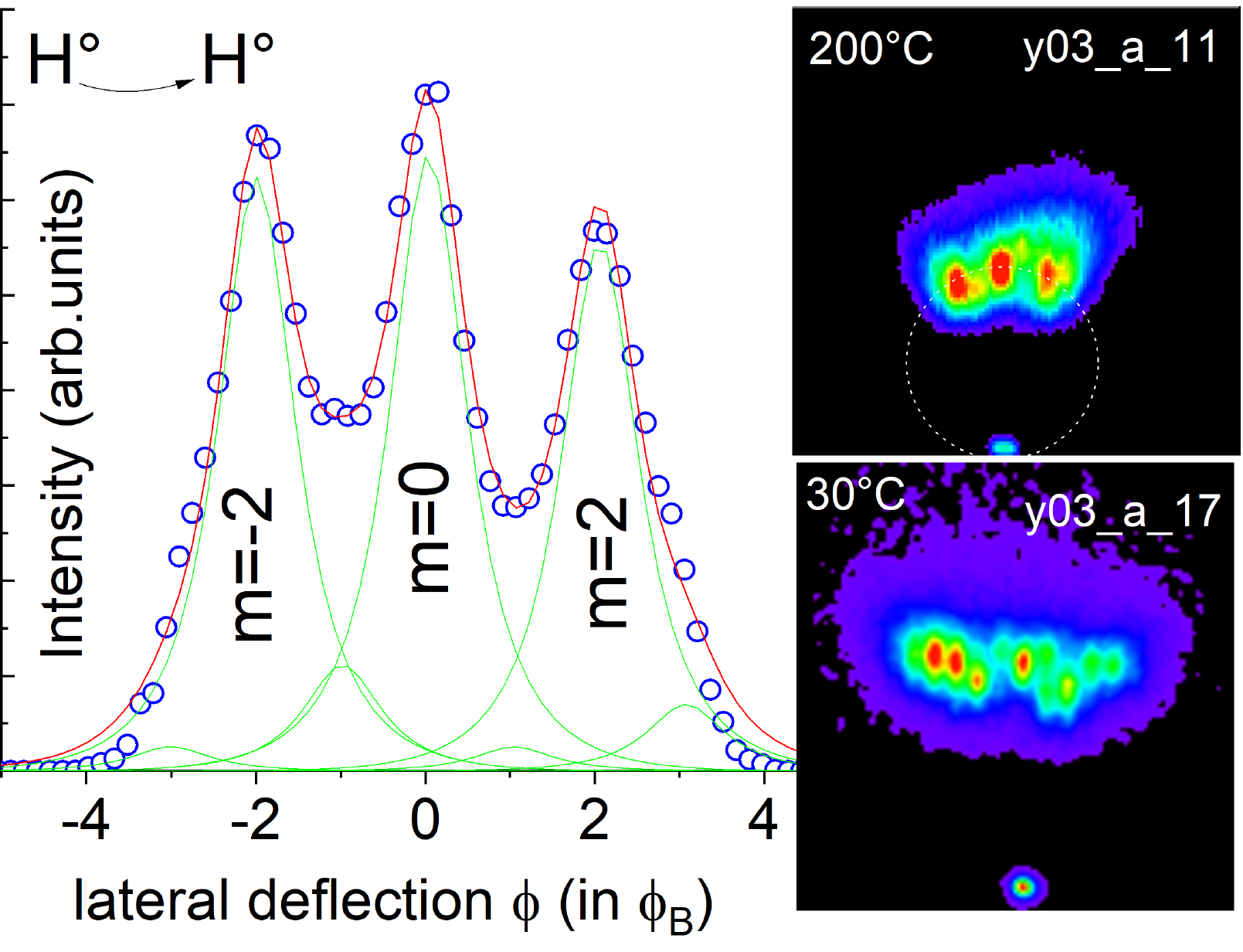}
\caption{\label{fgr:y03_a_11} First diffraction pattern of fast atoms on a surface recorded on Jan. 13$^{th}$ and 14$^{th}$ 2003. Here 500 eV H$^\circ$ projectiles on NaCl at 200$^\circ$C along $\langle 100 \rangle$. The angle of incidence is $\theta_i\sim$0.6$^\circ$ and $\phi_B$=0.146$^\circ$. The lower insert was recorded the day after at $\theta_i\sim$0.7$^\circ$ with a NaCl crystal at room temperature.}\end{figure}

Finally, Fig.\ref{fgr:y03_a_11} recorded with H$^\circ$ atoms shows the first evidence that something special was indeed taking place during the neutral part of the trajectory, with discrete spots equally separated by an angle $\phi_B$ along the $y$ direction.
The day after, the filament used to heat the target around 200 $^\circ$C broke and the same image with a target at room temperature displayed an even larger contrast.

Upon my return from my visiting committees, Patrick reported the finding and I asked him to wait a bit until I could check the detector electronics for a possible problem.
Everything became limpid during the visit of Fernando Mart\'{i}n who presented results from his collaboration with Daniel Far\'{i}as and Christina D\'{i}az on the diffraction of thermal H$_2$ molecules on a Pd surface\cite{farias_2004_pronounced}.
When it became clear that we were facing atomic diffraction Hocine Khemliche decided to apply for a patent\cite{Khemliche_patent} and the CNRS agreed to support us provided that we demonstrate that GIFAD also works with semi-conductors.
We were quite enthusiastic about this new challenge, Hocine, because we were getting closer to applications, while I could not wait to benefit from the quality of the surfaces produced by molecular beam epitaxy and agreed to postpone the publication.
\begin{figure}\centering\includegraphics[width=0.7\linewidth,draft = false]{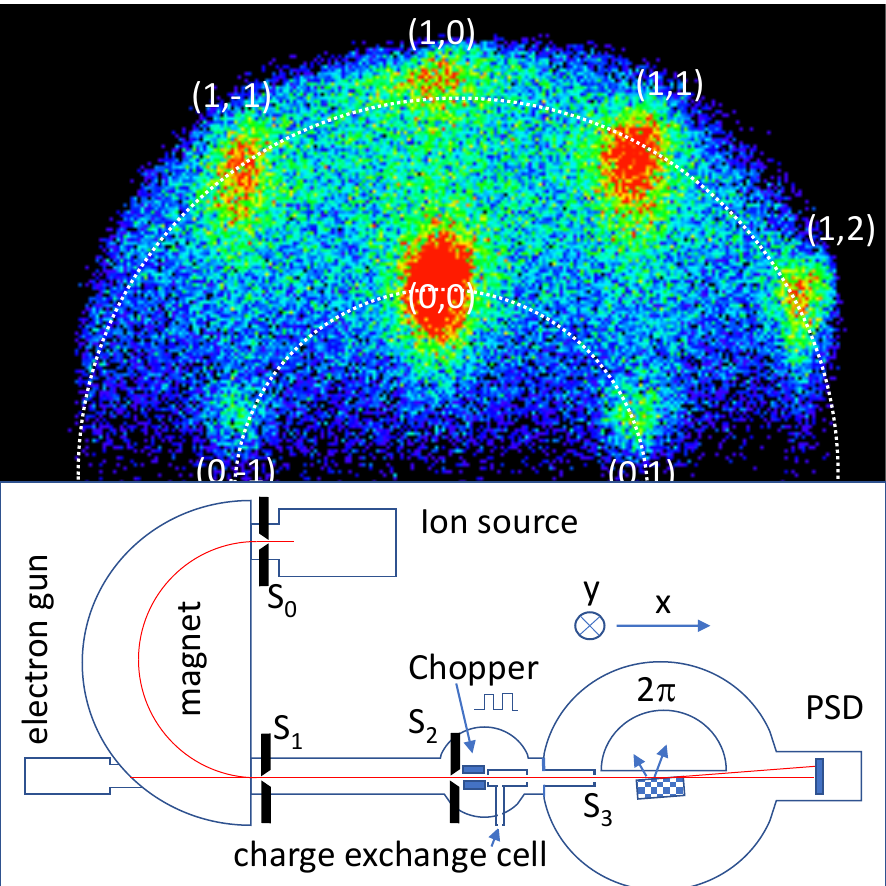}
\caption{\label{fgr:RHEED} RHEED image recorded \textit{in situ} in March 2006 on GaAs with a 30 kV electron gun taken out from an electron microscope (sold on eBay). The indexes correspond to the integers $(l,m)$ such that $\Vec{k_f} = \Vec{k_i} + l.\Vec{G_x} + m.\Vec{G_y}$ with $\Vec{G_x}, \Vec{G_y}$ the reciprocal lattice vectors. Two Laue circles are visible, the specular one ($l=0$) and an outer one ($l=1$). The inset is a scheme of the beamline \cite{Villette_these,Chopper}.}
\end{figure}
 
%To become familiar with RHEED, we managed to get an electron diffraction pattern by after I found on eBay a 30 kV electron gun from a electron microscope that Hocine went to collect in a barn in the country side. It was placed in the straight line available at the exit of the magnet installed for charge selection. Having neglected the importance of the earth magnetic field, we had to install magnetic coils taken from standard cathodic TV tubes to steer the electron beam on axis.

We soon realized that the MBE community is also close to paranoid about the idea of having any non-certified item inside an MBE chamber. 
We initiated a collaboration with the MBE group at Institut des Nanosciences de Paris (INSP) who were interested and had some knowledge on capping samples produced in UHV so that they could be exposed to air, transported, and then decapped in vacuum by thermal treatment.
Our first test with GaAs capped with antimony was unsuccessful because we had guessed that H$^\circ$ atoms would be the ideal projectile, neglecting the fact that H$^\circ$ atoms interact strongly with GaAs.
Switching to helium atoms we did not resolve the Bragg peaks and did not understand the peculiar polar profile made of a beautiful elastic peak on top of a very broad inelastic one!
We were more lucky with ZnSe(001) samples grown on GaAs(001) and capped with selenium prepared by Victor Etgens and Mahmoud Eddrief.
In February 2006, we could record nice diffraction profiles illustrating the power of GIFAD because it clearly indicates the charge transfer between Zn and Se atoms at the surface\cite{khemliche2009grazing}. We could proceed with the patent and develop a strategy to install GIFAD inside an MBE chamber.
The patent was accepted during the summer of 2006\cite{Khemliche_patent} so that I felt free to present results at IISC-16 in Schloss Hernstein, Austria.
We knew from a LEIF meeting in Denmark that H. Winter had obtained his first diffraction result.
We should have sent the paper to PRL before the conference because Helmut Winter was able to write it in the following week so that both contributions\cite{Schuller_2007,Rousseau_2007} were published back to back in the same volume but he submitted first.

Note that the possible diffraction of keV atoms had been predicted in 2002 in a Russian chemistry journal\cite{Andreev_2002} but attracted our attention only years later.

\section{FAD in Berlin}\label{ch:FAD_Berlin}
\begin{figure}\centering
\includegraphics[width=1.0\linewidth,draft = false]{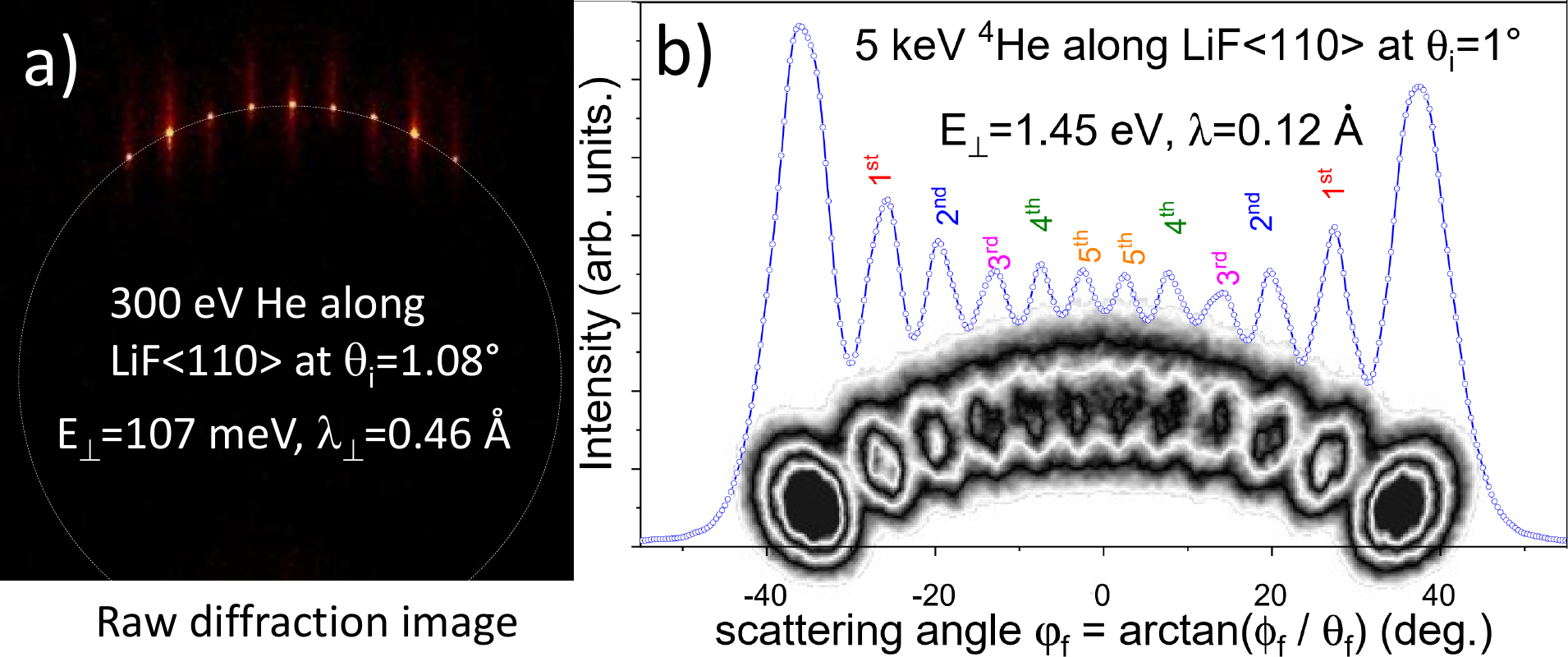}
\caption{\label{fgr:supernum} For helium scattering on LiF along the $\langle 110 \rangle$ direction. a) At a moderate value of $E_\perp$, sharp spots are visible located on the Laue circle, revealing a significant elastic diffraction. b) At larger values of $E_\perp$, no elastic peak is visible, and even inelastic one becomes difficult to isolate. Here, 32 diffraction orders can be identified but the overall azimuthal intensity modulation shows five supernumerary rainbows as originally described in Ref.\cite{Schueller2008Supernumerary} in addition to the outer classical rainbow.(Taken from Ref.\cite{debiossac_these}).}
\end{figure}

As far as I was told by Helmut Winter after checking in the laboratory notebook, Andreas Sch\"uller discovered fast atom diffraction in May 2006 when he was investigating interaction potential of the helium-LiF system via rainbow scattering\cite{schuller2005interatomic}.
Using neutral helium was a guarantee to remove any effect from the image charge.
In Berlin, the first hint was a surprising intensity modulation corresponding to unresolved diffraction peaks but revealing the supernumerary rainbow structure that they published immediately after\cite{Schueller2008Supernumerary} (in TEAS, supernumerary rainbows were described by Garibaldi \textit{et al.} \cite{garibaldi1975quantum} and Avrin and Merrill\cite{Avrin1994}).
I was not enthusiastic at first because the model is much less predictive than the hard-corrugated wall that can handle multiple valleys in the lattice unit.
However, I ended up being a bit jealous as the beauty of the optical rainbow is meaningful to most scientists and, historically the supernumerary rainbows triggered the unification of Descartes \cite{Descartes_1637} geometric description and Newton corpuscular description \cite{newton2014letter} with the wave description by Young theory of interference \cite{young1832bakerian} in the very beginning of the 19th century.
As with optics, the principal as well as the secondary rainbows in their simplest form \textit{i.e.} without the associated Airy profile, do not require a wavelength for their interpretation\cite{miret2012classical} but supernumerary rainbows do.
It is amusing to realize that Orsay and Berlin groups started with these two different manifestations of the wave nature of the keV projectiles illustrated in Fig.\ref{fgr:supernum}a) and Fig.\ref{fgr:supernum}b) respectively.
In both cases, the effect was not anticipated but position-sensitive detectors were present and waiting to measure the scattering profiles of light-neutral particles at crystal surfaces.

\section{A fierce competition}\label{Ch:competition}
Soon after, we received a significant grant from the French funding agency to adapt GIFAD inside a RIBER$^\circledR$ compact-21  to be installed in Paris at INSP.
With P. Rousseau, I started to develop a multiple collision description of the Debye-Waller factor adapted to GIFAD\cite{Rousseau_2008} while theorists from Vienna and Belgrade developed a random kick approach to the decoherence of the projectile wave-packet\cite{aigner2008suppression}.
H. Khemliche had the bright idea to invite J.R. Manson for a six-month visit to Orsay where he could teach us about various scattering regimes\cite{Manson_PRB_2008}.
We succeeded in observing the first diffraction on Ag (110), a metal surface\cite{bundaleski2008grazing} only shortly before Berlin published similar results on Ni(110) \cite{busch2009fast}.
Later, with N. Bundaleski and the advice of A.G. Borisov, we tried to understand why the absence of a bandgap and the associated electronic excitation do not destroy diffraction by modeling the electron scattering on the projectile at the Fermi edge\cite{khemliche2009electron,bundaleski2011decoherence}.
Berlin managed to observe diffraction from superstructures resulting from impurities embedded in an iron crystal that were brought to the surface by annealing \cite{schuller2009fast}.
They were the first to publish diffraction patterns recorded off-axis \cite{seifert2011transition} while A. Zugarramurdi and A.G. Borisov were the first to calculate it with a quantum code\cite{zugarramurdi_2012_transition,zugarramurdi2013surface}. With M. Debiossac, starting from the perturbation theory developed by C. Henkel\cite{Henkel_94} we derived an empirically enforced time-reversal symmetry to derive a formula for misaligned conditions in GIFAD\cite{Debiossac_PRA_2014}.
With B. Lalmi and P. Soulisse, we started to track mosaic domains \cite{Lalmi_2012}.
Berlin was the first to pay attention to the surface of perovskite \cite{dirsyte2009surface} trying to use GIFAD for online monitoring of surface improvement.
We managed to diffract on single-layer graphene on SiC(0001) while Berlin obtained results on single-layer silicon oxide on a Mo(112) surface\cite{Seifert_2009}. 
Berlin was the first to show that the sensitivity to the atomic position could be in the pm range\cite{schuller2010rumpling} and to observe the first breakdown of the axial channeling approximation (ASCA) when the atom direction is far away from a crystal axis\cite{busch_2012_evidence}, see also \cite{zugarramurdi_2012_transition} for a theoretical description.
We were first on semi-conductors\cite{khemliche2009grazing} inside a MBE vessel\cite{Debiossac_PRB_2014} and real-time tracking of the growth\cite{Atkinson_2014,Debiossac_2017} but Kyoto and Berlin had already demonstrated growth oscillations with keV ions\cite{fujii1993layer,igel1996intensity}.
We were probably the first to record simultaneously energy loss and diffraction pattern but a mistake in the tuning of the chopper made our initial interpretation erroneous\cite{erratum} so that the Berlin group was first to show that diffraction disappears completely when an electron is removed from the valence band\cite{lienemann_2011}.
They were the first to record diffraction on an organic layer with alanine on Cu(110) \cite{Seifert_2013} shortly before N. Kalashnyk and M. Debiossac also succeeded with perylene on Ag(110)\cite{Natalya}, a work that also triggered atomic triangulation \cite{Kalashnyk_2016} (see also Fig. 10 of Ref.\cite{pan2022setup}) which was also developed in Berlin\cite{feiten2015surface} in a negative-contrast where the peak amplitude is not quantitative.
They were the first to produce a nice comprehensive review on fast atom diffraction\cite{Winter_PSS_2011} with only one weak point, elastic diffraction was not yet clearly established\cite{khemliche2009grazing,busch_2012_evidence,Debiossac_PRB_2014} so there is no discussion on inelastic effects.
We were first to observe a bound state resonance on the surface\cite{debiossac2014transient,jardine_2004_ultrahigh}
and the Lamb-Dicke effect revealing the quantum nature of the surface atoms \cite{roncin2017elastic} opening the investigation of the inelastic profiles\cite{pan2021polar,pan2023lateral}.
I was a referee of some of the Berlin papers in PRL and a referee comment that our observation of bound state resonances\cite{debiossac2014transient} was a 'tour de force' suggests that Helmut Winter also was a referee of some of our papers.
We were both confident that despite the fierce competition, the evaluation would be fair.

\subsection{Other ion diffraction}
So far, the genuine diffraction of ions on surfaces has never been observed, the reason is probably that the electric field of the ions interacts strongly with the surface, via the real and imaginary part of the dielectric function.
The real part gives the image charge effect accelerating the ion towards the surface, increasing the impact energy ($E_\perp$) in the eV range while the imaginary part describes the direct coupling of the fast movement along $x$ to the optical phonon modes\cite{borisov1999evidence} turning the diffraction inelastic.
All these are significantly reduced if the ion neutralizes at comparatively large distance from the surface, as for the H$^+$-NaCl system above or H$^+$-LiF\cite{xiang_these}\cite{Rousseau_these} but also He$^+$ on iron, see \textit{e.g.} Fig.9 of Ref.\cite{schuller_NIMB_2009} by the Berlin group, which displays scattered intensities of He$^+$ and He$^\circ$ from 3 keV $^3$He$^+$ ions impinging on $c(1\times 3)$S/Fe(110) surface at 0.65$^\circ$. 
It shows well-resolved inelastic peaks at different exit polar angles indicating a complete neutralization on the way-in, at a large distance from the surface, and an image charge acceleration of $\sim$ 2 eV and well-defined supernumerary rainbows. 
However, none of these systems shows clear signs of an elastic diffraction peak.

\subsection{An old trick for large surface coherence}
In the grazing incidence community, surface preparation is a cumbersome but mandatory step where cycles of sputtering and annealing gradually remove impurities and improve the crystalline order of the surface.
On metals, the expertise of the Berlin group was so high that not many groups dared to compete.
When switching to ionic insulators, we focused on LiF samples with optical quality, \textit{i.e.} polished to $\lambda/4$, and indeed a gentle thermal treatment at a few hundred $^\circ$C is enough to get a decently flat surface with reflection coefficient close to unity at a few deg. incidence. However, so far, elastic diffraction has never been observed on these samples.
Before retiring, Marc Bernheim, co-organizer of IISC-9 in Aussois, offered me some of his equipment and introduced me to a colleague from the "Laboratoire de Physique du solide" who gave me a little plastic box with a few brown-orange LiF crystals that he had irradiated by neutrons some decades ago.
I did not pay too much attention until we accidentally damaged our LiF samples.
It is with these "old samples" that we recorded elastic diffraction in the form of sharp spots (see \textit{e.g.} Fig.\ref{fgr:supernum}a) allowing observation of bound state resonances.
Not only the contrast is much larger but the coherence of the scattered beam is mandatory to allow interference between directly reflected and temporarily trapped beams.
After a few weeks in a base pressure of a few 10$^{-10}$ mbar these elastic peaks progressively disappear and neither annealing nor sputtering could restore them.
From our limited understanding, the high density of color centers favors easy cleaving along atomic planes and a gentle annealing in UHV restores the transparent color.

\section{Acknowledgements}

We are indebted to the EEC for their continuous support in the FP3, FP5, FP6 programs under contract CHRT-CT93-0103, HPRI-CT-1999-40012 and RII3-026015.
We acknowledge funding from the Agence Nationale de la Recherche (Grants No. ANR-07-BLAN-0160-01 and ANR-2011-EMMA-003-01) as well as from the Triangle de la Physique (Grant No. 2012-040T-GIFAD).
We express our gratitude to all those who contributed to the ion-surface and atom-surface experiments in Orsay. A.G. Borisov, V.A. Morosov, A. Kalinin, J.Villette, Z. Szilagyi, J.P. Atanas, H. Khemliche, A. Momeni, Y. Xiang, N. Bundaleski, J.R. Manson, B. Lalmi, P. Soulisse, M. Debiossac, P. Lunca-Popa, P. Pan and our colleagues from molecular beam epitaxy in Paris V. Etgens, M. Eddrief, F. Finnochi and P. Atkinson.

 \bibliography{cas-refs}

\begin{thebibliography}{152}
\bibitem [{AIM()}]{AIM}%
  \BibitemOpen
  \href@noop {} {\bibinfo {title} {{\color{purple}We were also extremely lucky to be among the first users of the best ECR ion source at this time, the one developed in Grenoble and made available with a CEA-CNRS agreement }}}\BibitemShut {NoStop}%
\bibitem [{Ne2()}]{Ne2}%
  \BibitemOpen
  \href@noop {} {\bibinfo {title} {{\color{purple}Later, With A. Momeni and H. Khemliche, we decided to try again investigating multiply charged ions using the most simple He$^{2+}$ and Ne$^{2+}$ ions. We observed the 'bottle neck' paradox that more neutrals are scattered starting from Ne$^{2+}$ than from Ne$^{+}$ ions, but we could not identify the origin\cite{Momeni_these}.}}}\BibitemShut {Stop}%
\bibitem [{Nat()}]{Natalya}%
  \BibitemOpen
  \href@noop {} {\bibinfo {title} {{\color{purple}N. Kalashnyk was a Post-doc in our group where she explored several perylene/{Ag} deposition schemes until diffraction could be measured with the help of M. Debiossac finishing his PhD. The work was repeated and developped soon after on the same setup but published without any mention of their initial contribution.}}}\BibitemShut {Stop}%
\end{thebibliography}

\end{document}